\documentstyle[12pt]{article}
\newcommand{\beq}{\begin{equation}}
\newcommand{\eeq}{\end{equation}}

\begin{document}
\begin{titlepage}
\renewcommand{\thefootnote}{\fnsymbol{footnote}}

\begin{center} \Large
{\bf Theoretical Physics Institute}\\
{\bf University of Minnesota}
\end{center}
\vspace*{.3cm}
\begin{flushright}
TPI-MINN-95/10-T \\
UMN-TH-1340-95 \\
OUTP-95-14P \\
hep-th/9504141
\end{flushright}
\vspace{.3cm}
\begin{center} \Large
{\bf Two Phases of Supersymmetric Gluodynamics}
\end{center}
\vspace*{.3cm}
\begin{center} {\Large Ian I. Kogan} \\
\vspace{0.4cm}
{Theoretical Physics, 1
Keble Road,
 Oxford, OX1 3NP , UK} \\
\vspace{0.4cm}
 and \\
\vspace{0.4cm}
{\Large
Mikhail Shifman } \\
\vspace{0.4cm}
{\it  Theoretical Physics Institute, Univ. of Minnesota,
Minneapolis, MN 55455}\\
\vspace*{2cm}

{\Large{\bf Abstract}}
\end{center}

\vspace*{.2cm}
We argue that supersymmetric gluodynamics
has two phases with equivalent infrared behavior,
 one of which is asymptotically free and another one is superstrongly
 coupled in the ultraviolet domain.

\end{titlepage}

Supersymmetric (SUSY)  gauge theories are unique examples of
non-trivial four-dimensional theories where some dynamical aspects
are
exactly tractable. The first results of this type -- calculation of the
gluino condensate and the Gell-Mann-Low function -- were obtained
in the early eighties \cite{NSVZ1,NSVZ2}. The interest to the
miraculous features of the
supersymmetric theories was revived after the recent discovery
\cite{S1} -- \cite{S2} of a
rich spectrum of various dynamical scenarios that may be realized
with a special choice of the matter sector. The basic tools in
unraveling these scenarios are: (i) instanton-generated
superpotentials which may or may not lift degeneracies along
classically flat directions \cite{ADS}; (ii) the NSVZ $\beta$ functions;
(iii)
the property of holomorphy in certain parameters \cite{Shif1,Shif2};
(iv) various
general symmetry properties, i.e. the superconformal invariance at
the infrared fixed points and its consequences \cite{S2}.  A beautiful
phenomenon revealed in this way is the existence of a generalized
``electric-magnetic" duality in $N=2$ \cite{SE} and some versions of
$N=1$
theories \cite{S2}.

In this letter we use basically the same methods to argue the
existence of two interrelated phases in the supersymmetric
gluodynamics, the simplest SUSY gauge theory where we deal only
with the gluons and gluinos. Unlike the examples mentioned above in
this case parameters of the theory can not be adjusted in such a way
that a weak coupling regime is ensured in a certain limit providing
us with a clue to what happens in the strong coupling regime. Still
some conclusions are possible.

The Lagrangian of the theory has the form
\beq
{\cal L} = \frac{1}{g_0^2} W^2\mid_F +\mbox{h.c.}
\label{lagrangian}
\eeq
where $g_0$ is a complex parameter, the bare coupling constant at
the ultraviolet cut off $M_0$ (whose imaginary part is related to the
$\theta$ angle). Below we will use also $\alpha\equiv g^2/4\pi$.
The standard picture going with this Lagrangian is as follows:
the theory is asymptotically free which means that, given a fixed
correlation length $\Lambda^{-1}$, the coupling $g_0^2$ at the
ultraviolet cut off must be adjusted to be logarithmically small,
\beq
\alpha_0 \sim \frac{2\pi}{3T(G)}\,\, \frac{ 1}{\ln
(M_0/\Lambda)}
\label{AF}
\eeq
where $T(G)$ is the quadratic Casimir operator in the adjoint
representation (e.g. $T(G)=N$ for SU(N)). The theory is believed to be
confining in the infrared domain. The
choice $\alpha_0\ll1$ guarantees that the mass scale developing in
the infrared is small compared to $M_0$, $\Lambda\ll M_0$.

As will
be seen, the very same Lagrangian defines another phase of the
theory which is superstrongly coupled in the ultraviolet domain,
$\alpha_0 \gg 1$; still it flows to the same infrared limit as the
standard asymptotically free phase. In particular, the same
correlation length $\Lambda^{-1}$ is achieved provided that
\beq
 \alpha_0 \sim  \kappa(M_0/\Lambda )^3
\label{SSC}
\eeq
where
$$
\kappa =\frac{2\pi\mbox{e}}{T(G)}\, .
$$

Thus, a duality takes place -- the phases with a weak and a
superstrong couplings (SSC) at the ultraviolet cut off evolve towards
one
and the same infrared asymptotics.

Our starting point is the exact NSVZ $\beta$ function for SUSY
gluodynamics,
\beq
\beta (\alpha ) = -\frac{\alpha^2}{2\pi}\,\,  \frac{3T(G)}{1-
(T(G)\alpha
/2\pi )}\, .
\label{NSVZbeta}
\eeq
When these $\beta$ functions were originally derived
\cite{NSVZ2} it was meant
that they are exact to all orders in perturbation theory; the question
whether they are exact non-perturbatively was not addressed. Later
on it was shown that in some cases these $\beta$ functions do
acquire additional non-perturbative parts \cite{S3}. In SUSY
gluodynamics
(with no matter fields) Eq. (\ref{NSVZbeta}) is exact
non-perturbatively. The proof
was actually given in Ref. \cite{Shif3}. Let us reiterate the main steps
considering
for definiteness the SU(2) gauge group. Introduce two auxiliary
matter fields in the fundamental representation (1 flavor), with a
small mass term $m_0$. The theory is then in a weakly coupled
Higgs phase \cite{ADS}. The existence of a conserved U(1) current
{\em
unambiguously} dictates the form of the superpotential. It is crucial
that this form is saturated by one instanton, no multi-instanton
contributions are allowed.
The one-instanton
contribution above leads to the {\em exact} result for the gluino condensate
$\langle\lambda\lambda\rangle$. One then uses the property of the
holomorphy to exactly extrapolate this result to large values of
$m_0$. At $m_0\rightarrow\infty$ (or, more exactly, $M_0$) we
return back to SUSY gluodynamics, and the exact expression for
$\langle\lambda\lambda\rangle$ implies that the $\beta$ function
(\ref{NSVZbeta}) is exact in both, perturbative and non-perturbative
senses. (The one-instanton saturation of the superpotential,
precluding from getting non-perturbative corrections in the
$\beta$ function (\ref{NSVZbeta}), can be inferred from
\cite{Shif3} indirectly -- any terms other than the standard one-instanton
would destroy the
{\em exact} proportionality of $\langle\lambda\lambda\rangle$
to $\sqrt{m_0}$ established in \cite{Shif3}. A more general recent
analysis \cite{Nati} of all terms, perturbative and non-perturbative,
that can appear in the superpotential proves the very same fact directly.)

A peculiar feature of this $\beta$ function is that it changes sign at
$\alpha = 2\pi /T(G)\equiv \alpha_*$ not through zero, as it happens
at the
regular fixed points, but rather through pole. Nevertheless, changing
the sign results in the fact that $\alpha = 2\pi /T(G)$ is an infrared
attractive point, the theory flows to it in the infrared from both
sides,
the small and large $\alpha$ domains.  As a matter of fact, the
solution of the renormalization group equation for $\alpha$ is a
double-valued function of the normalization point $\mu$.  If one
starts from $\alpha_*$ at $\mu =\Lambda$ one solution evolves
to the standard asymptotically free phase, with a small value of
$\alpha$ at short distances, Eq. (\ref{AF}).  Another solution evolves
to a
superstrongly coupled (SSC) phase at short distances
(see Eq. (\ref{SSC}) and Fig. 1). Near the critical point
$$
\alpha = \alpha_*\pm\sqrt{6\alpha_*^2\Lambda^{-1}(\mu -
\Lambda )}\, .
$$
If
these two solutions are denoted by $\alpha_1$ and $\alpha_2$,
respectively, the effective infrared theory is invariant under the
interchange $\alpha_1\leftrightarrow\alpha_2$ at $M_0$. For large
$M_0$ this invariance reduces to
$$
\alpha_0 \rightarrow \kappa\left( {\rm e}^{\frac{2\pi}{T\alpha_0}}
-1\right)
+\frac{2\pi}{T}\frac{1}{\ln (\alpha_0/\kappa )}\, .
$$
If a lattice or a similar
formulation of the SUSY gluodynamics existed one could develop a
strong coupling expansion similar to Wilson's \cite{Wils}, and the
theory would  trivially confine color in this phase. The similarity
with Wilson's
argument ends quickly, though. Indeed the Wilson strong coupling
expansion assumes that the correlation length in the strong coupling
phase is of order of $M_0^{-1}$, while in our case the correlation
length is much larger, $\sim (M_0g_0^{-2/3})^{-1} \gg M_0^{-1}$.

If one studies only the vacuum condensates and other long distance
characteristics both phases are indistinguishable from each other.
Such quantities depend only on the Wilson coupling constant
\cite{Shif1},
$$
\alpha_W^{-1} =\alpha^{-1}-\frac{T}{2\pi}\ln\alpha^{-1}
$$
 which is the same in the  both phases in the
respective points $\alpha_{1,2}$. Incidentally, it is just the Wilson
coupling $\alpha_W^{-1}$ on which the chiral quantities and $F$
terms
depend holomorphically, not $\alpha^{-1}$. The distinction appears
when
one studies the short distance properties of different correlation
functions. Consider for instance, the two-point function
\beq
\langle T\{ \lambda (x) \lambda (x) \, , \,\,\, \bar\lambda (0)
\bar\lambda (0)\}\rangle\, .
\label{CF}
\eeq
In the asymptotically free phase at $|x|\ll\Lambda$ it behaves as
$x^{-6}(\ln x)^{-2}$ and is very singular at $x\rightarrow 0$.
The corresponding spectral density at large $s$ grows linearly
with $s$. In the SSC phase, where the interaction becomes strong at short
distances, the bound states (except for a few low-lying ones) are
generically expected to have sizes of order $M_0^{-1}$.
One can expect something similar ``fall to the center". Then the spectral
density must fall off rapidly above the lowest state (which is the same in
both phases). Accordingly, the singularity at small $x$ in the correlation
function (\ref{CF}) is much softer in the SSC phase.

We  would like to add a speculative remark on  the nature of the
 point $\alpha_{*}$ which may be a bifurcation point of
 the renormalization group (RG)  flow.  Formally the running
coupling constant corresponding to the     $\beta$ function
 (\ref{NSVZbeta}) satisfies the relation
\beq
\frac{\kappa}{\alpha(\mu) }\exp\left[-
 \frac{2\pi}{\alpha(\mu) T}\right] =
 \left(\frac{\Lambda}{\mu}\right)^{3}\, .
\label{running}
\eeq
It is easy to see that Eq. (\ref{running}) has no real solutions for $\mu <
\Lambda$. However, one can find a complex solution. For example
 if $\mu$ is slightly smaller than $\Lambda$, i.e.
$(\Lambda/\mu)^{3} = (1+\epsilon)$ (where
 $\epsilon \ll 1$) one can continue the RG trajectories
$$
\frac{1}{\alpha(\mu) } =\frac{1}{\alpha_*}\left[\left(1-
\frac{2}{3}\epsilon\right)
 \pm i\sqrt{2\epsilon}\right]\, .
$$
This looks like  generation of a $\theta$ term,
$
{2\pi}/{\alpha } \rightarrow {2\pi}/{\alpha } + i\theta
$, which is unobservable, anyway.

 Such a behavior is rather unusual in  quantum
 field theories, but as
 an example of such a
strange behavior
 let us  remind the reader about a peculiar  RG flow in the
two-dimensional O(3)
$\sigma$ model with the $\theta$ term \cite{Aff}, i.e.
 in the theory with the action
$$
S = \frac{1}{2g^{2}}\int d^2 x
\partial_{\mu}\vec{n}\partial_{\mu}\vec{n}
 + i \frac{\theta}{8\pi}
\int d^2 x \epsilon_{\mu\nu}\left(\vec{n}\left[
\partial_{\mu}\vec{n} \times\partial_{\nu}\vec{n}\right]\right)\, .
$$
 The RG flow here is as shown on Fig. 2;  one can see that at $\theta =
\pi$
 there is an unstable fixed point ($\beta$ function for $g$
 has a double zero at some $g_{*}$) which  can be considered either
 as an infrared fixed point for the  trajectories  starting at small $g$
 and $\theta = \pi$ or as an ultraviolet fixed point  leading to strong
 coupling phase with all possible values for $\theta$. One can observe
 some analogy between this behavior and the $\alpha_{*}$ point
 arising because of the pole in the NSVZ  $\beta$ function. To reveal
this analogy it is more appropriate to consider the  zero charge case
rather than the asymptotically free theory.  This is achieved by
adding $N_f$ massless matter flavors ($2N_f$ chiral superfields). The
$\beta$ function in
 this case is (for the gauge group SU(N))
\beq
\beta (\alpha ) = -\frac{\alpha^2}{2\pi}\,\,  \frac{3N-
N_{f}T(1-\gamma(\alpha))}
{1- (N\alpha
/2\pi )}\, .
\label{NSVZbetamatter}
\eeq
where $\gamma(\alpha)$ is the anomalous dimension of the  matter
superfields
 in the fundamental representation.  For $N_{f}$  large enough
 the
 weak coupling phase  is  actually infrared free (the Landau zero
charge), and the pole point $\alpha_*$   becomes an ultraviolet
attractor unless the behavior of $\gamma (\alpha )$ screens it
(i.e. the numerator of Eq. (\ref{NSVZbetamatter}) develops zero before the
denominator). In other words, we assume that, unlike the situation in Ref.
\cite{S2},  the conformal point does not develop at $N_f>3N$.
Then again,  the critical value $\alpha_*$ is achieved
at a finite value of the normalization point; the question arises as to
what happens when $\mu$ evolves further, to higher values.
  In this context the problem is even more striking --
 if in the case of the asymptotically free theory one can insist that
because of confinement one simply can not  go below $\Lambda$,
 in the latter case we {\em have} to define the theory in the ultraviolet
limit.
Presumably one can
 think about
this  further evolution  along the lines suggested by the $\sigma$
model above, i.e. considering this point as a bifurcation of the RG flow. This
question deserves
 further investigation.

\vspace{1cm}

{\bf Acknowledgments} \vskip .2in \noindent
We thank A. Rosly and A. Vainshtein   for discussions.
 I. K. is  grateful to  all members of TPI for hospitality  during his
visit in
  March-April 1995. This work was supported
in part by   DOE under the grant number
 DE-FG02-94ER40823, PPARC grant GR/J 21354 and by
   Balliol College, University of Oxford.

\newpage
\begin{center}
{\Large Figure Captions}
\end{center}
\vspace{ 1cm}

Fig. 1.\\
The double-valued solution of the renormalization group equation for the
running coupling constant in supersymmetric gluodynamics. \\

\vspace{ 1cm}
Fig. 2. \\
The renormalization group flow in the two-dimensional O(3) $\sigma$ model
with the $\theta$ term.

\newpage



\vspace{0.5cm}

\begin{thebibliography}{99}

\bibitem{NSVZ1}
V. Novikov, M. Shifman, A. Vainshtein and V. Zakharov,
{\it Nucl. Phys.} {\bf B229} (1983) 407.

\bibitem{NSVZ2}
V. Novikov, M. Shifman, A. Vainshtein and V. Zakharov,
{\it Nucl. Phys.} {\bf B229} (1983) 381; {\it Phys. Lett.}
{\bf B166} (1986) 329 (The  Gell-Mann-Low functions of
the type presented in Eqs. (\ref{NSVZbeta}),
(\ref{NSVZbetamatter})
will be referred to below as the NSVZ $\beta$ functions).

\bibitem{S1}
N. Seiberg, {\it Phys. Rev.} {\bf D 49} (1994) 6857.

\bibitem{ILS}
K. Intriligator, R. Leigh and N. Seiberg, {\it Phys. Rev.} {\bf D 50}
(1994) 1092.

\bibitem{IS}
K. Intriligator and N. Seiberg, {\it Nucl. Phys.} {\bf B431} (1994) 551.

\bibitem{S2}
N. Seiberg, {\it Nucl. Phys.} {\bf B435} (1995) 129.

\bibitem{ADS}
I. Affleck, M. Dine and N. Seiberg,
{\it Nucl. Phys.} {\bf B241} (1984) 493;  {\bf B256} (1985) 557.

\bibitem{Shif1}
M. Shifman and A. Vainshtein,
{\it Nucl. Phys.} {\bf B277} (1986) 456.

\bibitem{Shif2}
M. Shifman and A. Vainshtein,
{\it Nucl. Phys.} {\bf B359} (1991) 571.

\bibitem{SE}
N. Seiberg and E. Witten, {\it Nucl. Phys.} {\bf B426} (1994) 12;
(E) {\bf B430} (1994) 485;  {\bf B431} (1994) 484.

\bibitem{S3}
N. Seiberg, {\it  Phys. Lett.} {\bf 206B} (1988) 75.

\bibitem{Shif3}
M. Shifman and A. Vainshtein,
{\it Nucl. Phys.} {\bf B296} (1988) 445.

\bibitem{Nati}
N. Seiberg, {\it Phys. Lett.} {\bf B318} (1993) 469.

\bibitem{Wils}
K. Wilson, {\it Phys. Rev.} {\bf D10} (1974) 2445;
{\it Phys. Rep.} {\bf 23} (1976) 331.

\bibitem{Aff}
I. Affleck, in {\it Fields, Strings and Critical Phenomena},
 eds. E. Brezin and J. Zinn-Justin (North-Holland, 1990),
 p. 563, and references therein.

\end{thebibliography}
\end{document}